\documentclass[preprint,onecolumn]{revtex4}

\usepackage{amssymb,amsfonts,amsmath,amsthm}
\usepackage{color,enumerate,graphicx}
\usepackage{hyperref}
\usepackage[normalem]{ulem}

\begin{document}

\title{Electric field control of interfacial Dzyaloshinskii-Moriya interaction  in Pt/Co/AlO$_x$ thin films}

\author{Marine Schott}
\author{Laurent Ranno}
\affiliation{ Univ. Grenoble Alpes, CNRS, Grenoble INP, Institut Néel, 38000 Grenoble, France}
\author{H\'el\`{e}ne B\'ea}
\author{Claire Baraduc}
\author{St\'ephane Auffret}
\affiliation{ Univ. Grenoble Alpes, CEA, CNRS, SPINTEC, F-38000, Grenoble, France}
\author{Anne Bernand-Mantel}
\email{bernandm@insa-toulouse.fr}
\affiliation{Universit\'e de Toulouse, Laboratoire de Physique et
  Chimie des Nano-Objets, UMR 5215 INSA, CNRS, UPS, 135 Avenue de
  Rangueil, F-31077 Toulouse Cedex 4, France}

\begin{abstract}
We studied  electric field modification of magnetic properties in a Pt/Co/AlO$_x$ trilayer via magneto-optical Kerr microscopy.  We observed the spontaneous formation of labyrinthine magnetic domain structure due to thermally activated domain nucleation and propagation under zero applied magnetic field. A variation of the period of the labyrinthine structure 
under electric field is observed as well as saturation magnetization and magnetic anisotropy variations. Using an analytical formula of the stripe equilibrium width  we estimate the variation of the interfacial Dzyaloshinskii-Moriya interaction under electric field as function of the exchange stiffness constant. 
\end{abstract}

\maketitle
\section{\label{sec:intro}Introduction}

 Electric-field (EF) control of magnetism in metals is currently an actively expanding field of study, motivated by potential applications in low power spintronics devices. The magnetization of metallic magnetic materials in spintronics devices can be manipulated using magnetic field or spin-polarized currents flow. The ability to control the magnetic properties of a system using a gate voltage is a promising technique which could lead to the development of EF-assisted magnetization switching devices. Since the first experimental demonstration of EF control of coercive field in a FePt thin film by Weisheit et al. 
  \cite{Weisheit349}, lots of studies have been conducted with the final intention of finding the best conditions for functional devices.
Significant variations in the magnetic anisotropy energy under EF application have been observed by many groups \cite{group:19782:Maruyama2009,group:19782:Wang2012,Niranjan10,group:19782:PhysRevLett.102.187201,group:19782:endo:212503,group:19782:1882-0786-6-7-073004}, as well as EF variation of $T_c$ in ultrathin films \cite{citeulike:14357902,group:19782:Chiba2011,group:19782::/content/aip/journal/apl/100/12/10.1063/1.3695160}. Recently, the EF impact on the exchange stiffness parameter $A_\mathrm{ex}$ has been addressed experimentally \cite{ando16,dohi16} and theoretically \cite{oba15}. In these systems, a dielectric oxide layer is present at the top surface of the ferromagnet to allow the application of an EF in a capacitor geometry. This induces a broken spatial inversion symmetry for the ferromagnetic thin film. 
The presence of these non-identical interfaces is the source of an antisymmetric type of exchange, the Dzyaloshinskii-Moriya interaction (DMI), which has been shown to be a source of chiral spin textures in Pt/Co/AlO$_x$ for instance \cite{belmeguenai15}. 
The interfacial origin of this DMI and its consequences in spin textures in ultrathin films make DMI another interesting EF tunable parameter. However, distinguishing EF modulation on DMI from other contributions is not straightforward as direct access to DMI is difficult.
In addition to theoretical predictions for ultra-thin samples \cite{Yang2018}, several experimental studies have reported a change in the DMI factor under EF application  \cite{Yang2018,citeulike:13767084,Srivastava2018,Koyama2018,Zhang2018,Suwardy2019}. This control was shown for relatively thick layers of Fe (20 nm) \cite{citeulike:13767084,Suwardy2019}, or a material presenting a weak DMI, Ta/FeCoB/TaO$_x$, as demonstrated by Srivastava et al. \cite{Srivastava2018}. 
More recently, Zhang et al. \cite{Zhang2018} and Koyama et al. \cite{Koyama2018} showed larger EF induced variation of DMI in Pt/Fe/MgO and Pt/Co/Pd/MgO systems presenting intermediate DMI values. \\
In this work, we analyzed labyrinthine domains (also referred to as stripe domains) in samples of Pt/Co/AlO$_x$ presenting large DMI values \cite{belmeguenai15,vanatka15}. 
A reversible evolution of the labyrinthine magnetic domain configurations under EF is observed. We also report an EF variation of the saturation magnetization and anisotropy field. The analysis of the variation of equilibrium stripe width 
allowed us to estimate the EF variation of the DMI term $D$. We stress here that this estimation strongly depends on the assumptions made on the exchange constant value.\\

\section{\label{sec:sys}System description and characterization}
\begin{figure*}
\begin{center}
\includegraphics[keepaspectratio=true,width=01.00\columnwidth]{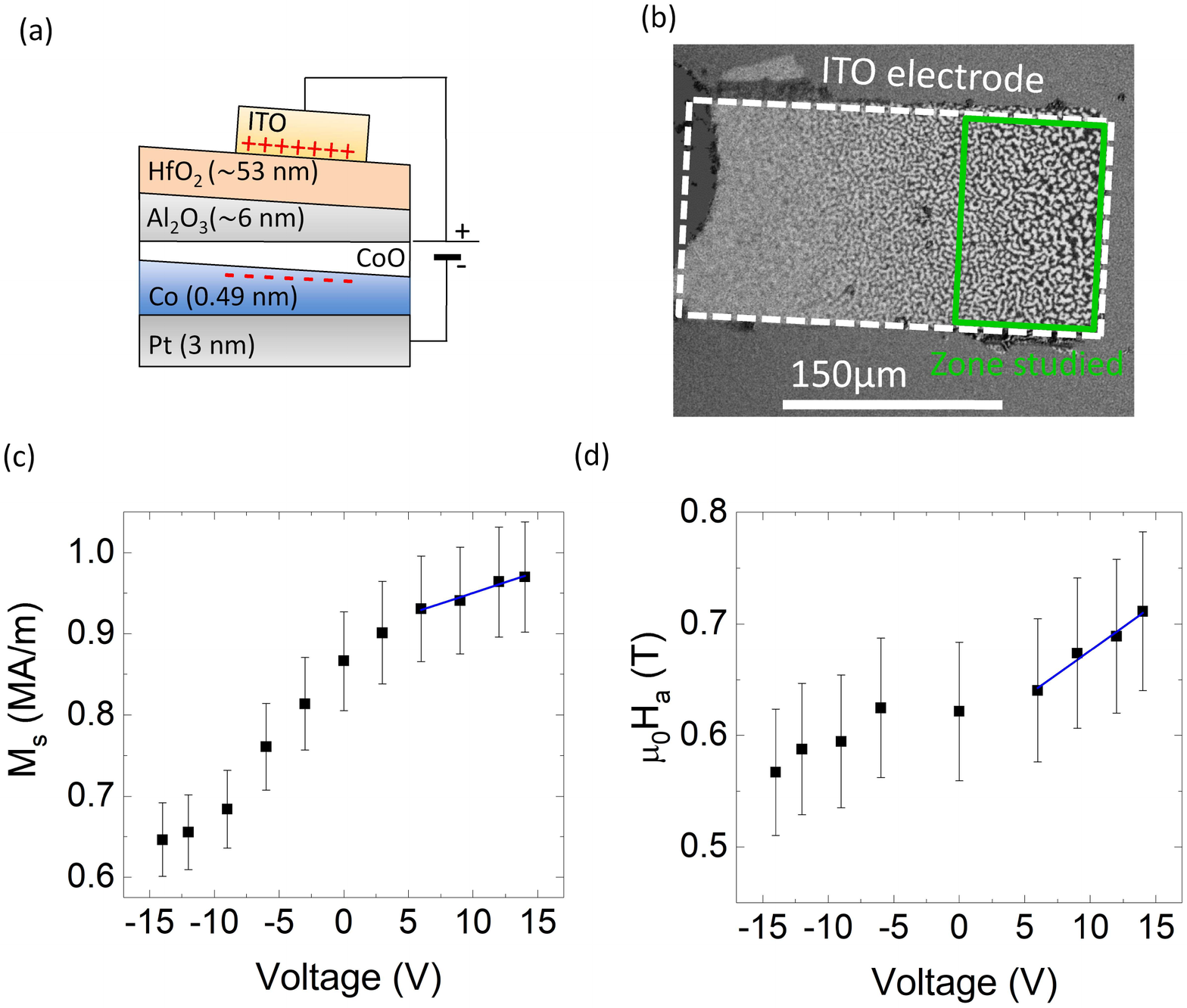}
\end{center}
\caption{(a) Schematics of the wedge sample and of the set-up used for EF application: the Pt/Co/AlO$_x$ trilayer is covered by a $\sim$ 50 nm  of dielectric layer (HfO$_2$) and a transparent top electrode (ITO). 
(b) p-MOKE image of the demagnetized state taken at   0 T, after saturation with an out-of-plane field of 0.4 mT for +12 V applied gate voltage. (c) Evolution of saturation magnetization with gate voltage. A systematic error of 7\%, was estimated from the error on the VSM-SQUID measurement. (d) Evolution of anisotropy field with voltage measured from p-MOKE hysteresis loops with in-plane field.  A systematic error of 10\%, was estimated from fluctuations in the field calibration of the p-MOKE measurement with in-plane applied magnetic field. The blue solid lines in c and d show the regions where the EF-efficiencies have been extracted. }
\label{fig:set-up}
\end{figure*}

Our system is a Pt(3 nm)/Co(0.6 nm)/AlO$_x$ sputtered trilayer, presenting a gradient of oxidation at the Co/AlO$_x$ interface. This was induced by the post-oxidation of a wedged-shaped Al top layer (Fig.~\ref{fig:set-up}(a)). Using this technique, we get access to a low thicknesses range for Co due to the Co partial oxidation in the region where the deposited Al is thin. The sample has been covered by about 50 nm of a high-$k$ insulator, HfO$_2$, deposited using atomic layer deposition. Then an Indium Tin Oxide (ITO) layer was DC sputtered and patterned in electrodes. These transparent electrodes allow  to obtain magneto-optical Kerr effect (MOKE) images of the magnetic domains in the Co layer under a gate voltage (Fig.~\ref{fig:set-up}(b)). Voltages of different magnitudes were applied between this top electrode and the Co layer. Here a positive voltage application induces electrons accumulation at the surface of the Co layer (Fig.~\ref{fig:set-up}(a)).
We used MOKE in polar geometry and recorded images and magnetic hysteresis loops on a single position on the sample through the ITO electrode under different applied voltages (Fig.~\ref{fig:set-up}(b)). 
To determine the EF variation of the saturation magnetization $M_\mathrm{s}$, we measured the variation of the p-MOKE signal between the two opposite saturated states, for out-of-plane applied magnetic field. The average value of magnetization of the wedged sample has been obtained by measuring $M_s$ with a Vibrating Sample Magnetometer-Superconducting Quantum Interference Device (VSM-SQUID). A $T_c$ of 366 K was estimated from VSM-SQUID measurements on  the sample under study  which possesses an averaged Co thickness of 0.49 nm due to partial oxidation of the deposited Co (see Fig.~\ref{fig:set-up}(a)).\\ 
 A strong EF dependence of $M_s$ is shown in Fig.~\ref{fig:set-up}(c). Such sensitivity of the saturation magnetization is due to the proximity of the $T_c$ to the measurement temperature (room temperature). Similar EF control of  $T_c$   associated with  strong $M_s$ variations were observed in recent studies \cite{ando16,chiba13}. 
The EF variation of the anisotropy field $\mu_0 H_a$ is presented in Fig.~\ref{fig:set-up}(d). The anisotropy field $\mu_0 H_a$ is extracted using Stoner-Wohlfarth  fit of the magnetization rotation toward the hard axis direction, on p-MOKE hysteresis loops, for in-plane applied magnetic field. 
From the measurements of $\mu_0 H_a(V)$, we deduced the variations of the effective magnetic anisotropy energy from the formula $\mu_0 H_a= 2 K_{\mathit{eff}}/M_s$, where  $K_{\mathit{eff}}=K_s/t-\frac{1}{2} \mu_0 M_s^{2}$ is the effective magnetic anisotropy and $K_s$ is  the surface magneto cristalline anisotropy. A voltage controlled magnetic anisotropy  (VCMA) coefficient   \cite{dieny17}  $\beta_{\mathit{PMA}}= \Delta K_s/\Delta E=$ 562 $\pm$ 102 fJ/(V$\cdot$m) is obtained. This is larger compared to the usual range obtained for a charge accumulation effect (10-290 fJ/(V$\cdot$m)) \cite{group:19782:Wang2012,group:19782:endo:212503,group:19782:Kita2012,PhysRevApplied.5.044006}. The enhanced effect is again likely due to the proximity of   $T_c$ to the measurement temperature. 

\section{\label{sec:leq}EF variation of equilibrium labyrinthine domains}
\begin{figure*}[h]
\begin{center}
\includegraphics[keepaspectratio=true,width=01.0\columnwidth]{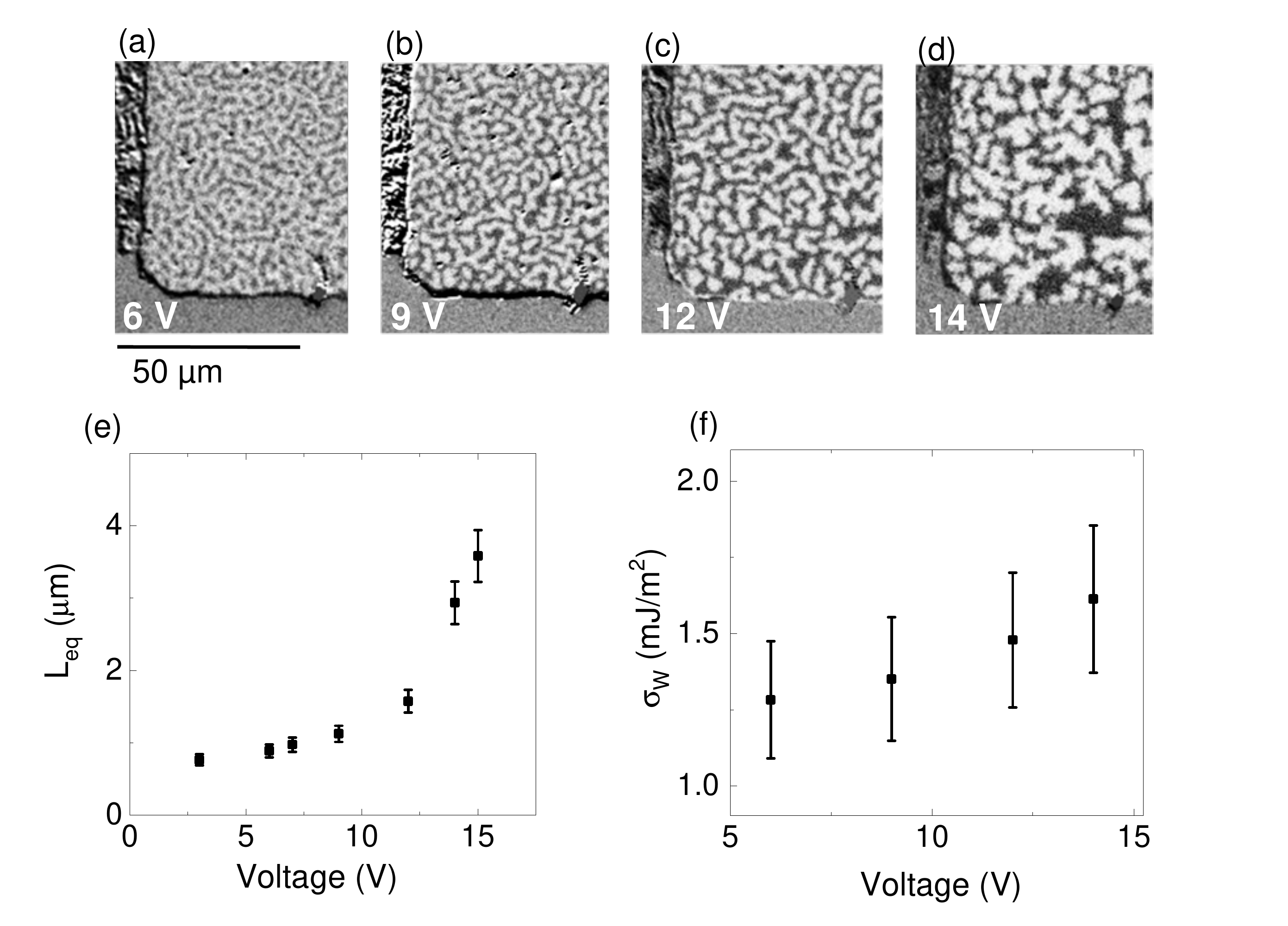}
\end{center}
\caption{(a)-(d) p-MOKE images of the evolution of equilibrium stripe width $L_{eq}$ with gate voltage. (e) Equilibrium stripe width $L_{eq}$ as a function of gate voltage.  We estimated from repeated measurements a $\sim$ 10\% error on $L_{eq}$ due to imperfect demagnetization. (f) Deduced domain wall energy as function of gate voltage. The error bars correspond to a 16\%  error calculated from the propagation of the errors in $M_s$, $t$ and $L_{eq}$.}
\label{fig:Leq}
\end{figure*}

As discussed earlier, the analysis of the equilibrium domain configuration has been shown to be one of the ways to study the influence of the EF on  different magnetic properties \cite{ando16,dohi16}. 
At zero and negative gate voltages, the domains are blurry due to a very strong thermally induced domain wall motion, as observed in previous works \cite{Bergeard2012}. Thus, in the following we focus our study on positive gate voltages,  for which the labyrinthine domains are stable, between 6 V and 14 V. To record the images of labyrinthine states and their variation with the EF, we proceed as follow: under a constant EF, we first saturate the sample with an applied magnetic field, then we turn off the magnetic field and record an image after a waiting time of a few seconds. The images presented in Fig.~\ref{fig:Leq}(a)-(d) correspond to the labyrinthine domains which spontaneously appear due to thermally activated nucleation and domain wall propagation. The labyrinthine domains are clearly and reversibly influenced by the application of an EF. The characteristic labyrinthine domain width $L_{eq}$ is estimated using a 2D Fourier transform of the images in Fig.~\ref{fig:Leq}(a)-(d).  In Fig.~\ref{fig:Leq}(e) we present $L_{eq}$ as function of the applied voltage, which shows an increase as a positive voltage is applied to the top ITO electrode. 
To understand in which way the EF is influencing the domain periodicity, we used an analytical model to describe the domain width $L_{eq}$ in ultrathin films:
\begin{equation}
L_{eq} =C\cdot t \cdot exp\left(\frac{\pi \sigma_\omega}{2K_dt}\right),
\label{width} 
\end{equation}
where $t$ is the cobalt thickness (which is fixed here to 0.49 nm), $K_d$ the dipolar constant $K_d=\frac{1}{2} \mu_0 M_s^{2}$ and $C$ is a model-dependent constant \cite{group:20379:KAPLAN1993111,GEHANNO199726} of order unity. The experimental domain wall energy $\sigma_\omega$ in Fig.~\ref{fig:Leq}(f) is deduced using the measured   $L_{eq}$ and $M_s$ values. We find domain wall energies $\sigma_\omega$ around $1.3-1.6$ mJ/m$^{2}$ for the range of voltages under study, which is consistent with earlier studies  \cite{doi:10.1021/acs.nanolett.7b00328}. \\

\section{\label{sec:dmi}EF variation of DMI}
\begin{figure}[h]
\begin{center}
\includegraphics[keepaspectratio=true,width=01.1\columnwidth]{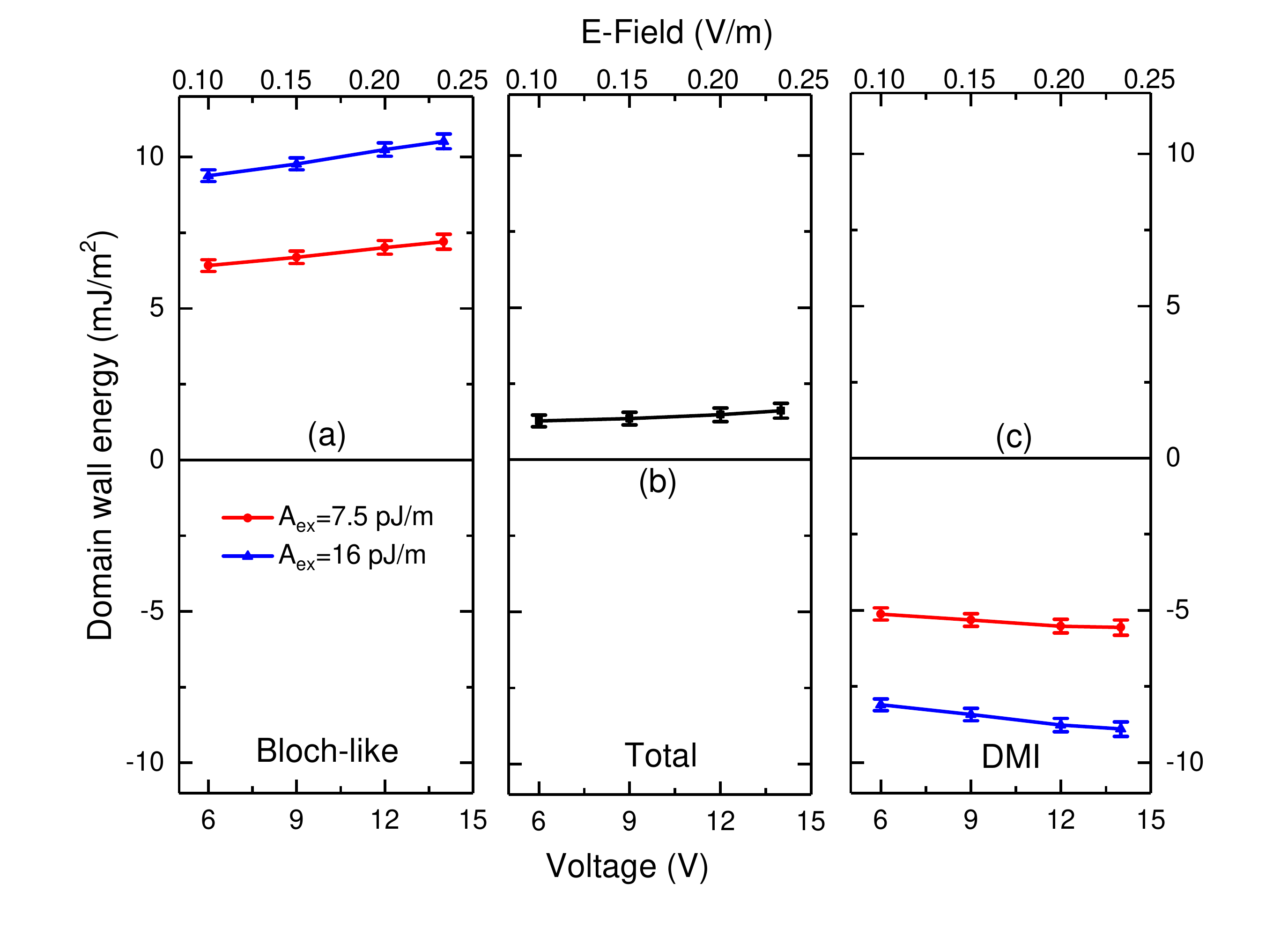}
\end{center}
\caption{Different components of the wall energy  $\sigma_w$ as a function of applied EF : (a) Bloch wall energy $\sigma^{Bloch}_w$. (b) Total energy $\sigma_w$.   (c)   DMI contribution $\sigma^{DMI}_w$.}
\label{fig:DMI}
\end{figure}
In the presence of sufficiently strong DMI, the domain wall is of N\'eel type with a sense of rotation which lowers the domain wall energy \cite{PhysRevB.78.140403}: $\sigma_ w= \sigma^{Bloch}_w+ \sigma^{DMI}_w =4\sqrt{A_{\mathit{ex}}K_{\mathit{eff}}} - \pi D$, where $\mathit{D}=D_s/t$ stands for the bulk DMI constant and is expressed in J/m$^2$, while the surface DMI value $D_s$ is expressed in J/m. We see that in order to extract $\mathit{D}$ using the measured $\sigma_w$ and $K_{eff}$ values it is necessary to know the exchange constant $A_{\mathit{ex}}$. We have used two values for $A_{\mathit{ex}}$. The first value $A_{\mathit{ex}}$=7.5 pJ/m (for V=0 V) was deduced from a fit using Kuzmin formula \cite{Kuz} of the temperature dependence of $M_s$ in our sample measured by VSM-SQUID.  This reduced value is coherent with recent studies \cite{Yastremsky2019}.
The second value $A_{\mathit{ex}}$=16 pJ/m (for V=0 V) is a value corresponding to bulk thin Co films \cite{PhysRevLett.99.217208}. The fact that the value of $A_{\mathit{ex}}$ extracted from our VSM-SQUID measurements is  lower than the bulk Co value is linked to the low $T_c$ of our sample (366 K) and coherent with the low $M_s$ value. For the same reason 
we can expect $A_{\mathit{ex}}$ to be influenced by EF as it was already suggested by previous studies \cite{ando16,oba15,dohi16}. 
We present the different deduced energy terms and their EF variation in Fig.~\ref{fig:DMI}. The total domain wall energy $\sigma_w$, also shown in Fig.~\ref{fig:Leq}(f) is presented in Fig.~\ref{fig:DMI}(b). In Fig.~\ref{fig:DMI}(a) we present two values and variations of the  Bloch wall energy
$\sigma^{Bloch}_\omega= 4\sqrt{A_{\mathit{ex}}K_{\mathit{eff}}}$ deduced from the measured $H_a$ and $M_s$ and the two values of $A_{\mathit{ex}}$ discussed ealier. $A_{\mathit{ex}}$ is considered to vary with EF as $\Delta A_{\mathit{ex}}/A_{\mathit{ex}} =2 \Delta M_s/M_s$, as the general tendency described by mean field approximation is that $A_{\mathit{ex}} \propto M_s^{2}$ near $T_c$ \cite{PhysRevB.82.134440}. Other theoretical studies have found a scaling law of $A_{\mathit{ex}} \propto M_s^{2-\epsilon}$, with $\epsilon$ being close to  0.2 for low temperatures  \cite{PhysRevB.94.104433,PhysRevB.92.134422,PSSB:PSSB200301671}. Fixing $\Delta A_{\mathit{ex}}/A_{\mathit{ex}} =2 \Delta M_s/M_s$ is then a good approximation and will give us an upper limit for the variation we can expect for $D(V)$. 
From these assumptions we deduce the DMI value and EF variations which are presented in Fig.~\ref{fig:DMI}. The $D$ value at 6 V is ranging between 1.3-2.4 mJ/m$^2$ depending on the chosen $A_{\mathit{ex}}$ value. We see that EF variation of $\sigma^{Bloch}_w$ (Fig.~\ref{fig:DMI}(a)) is stronger than the variation of the total wall energy $\sigma_w$ (Fig.~\ref{fig:DMI}(b)) and that an EF variation of $D$ is necessary to compensate this variation (Fig.~\ref{fig:DMI}(c)).
\begin{figure}[h]
\begin{center}
\includegraphics[keepaspectratio=true,width=00.8\columnwidth]{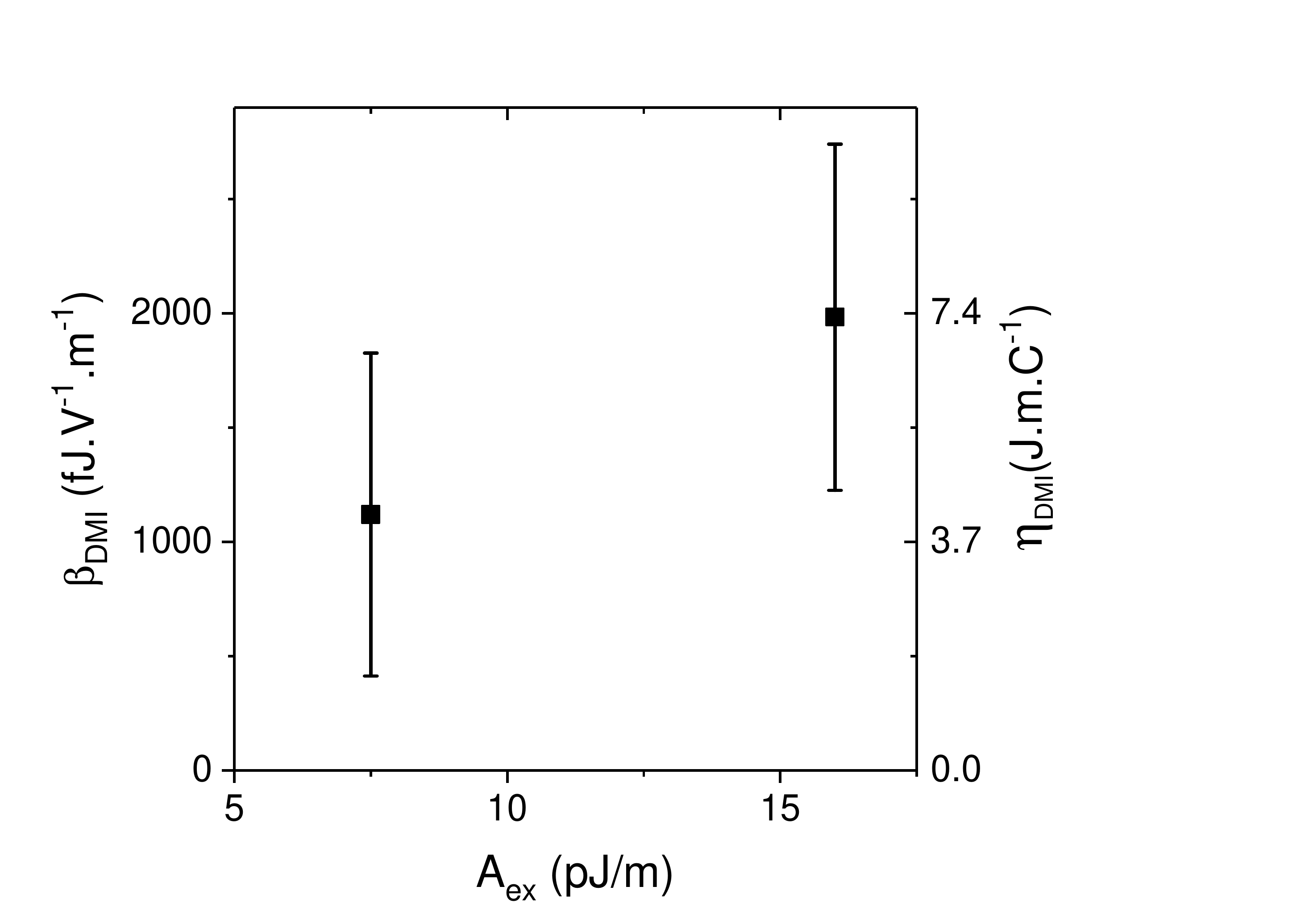}
\end{center}
\caption{ $\beta_{\mathit{DMI}}$ coefficients deduced from the experimental $M_s$, $t$, $L_{eq}$ and $K_{\mathit{eff}}$ values as a function of the chosen $A_{\mathit{ex}}$ value. The error bars are calculated from the standard error on the slope on a linear fit of $D(V)$.}
\label{fig:BDMI}
\end{figure}

\begin{table}[!ht]

\footnotesize{
   \begin{tabular}{|@{\hskip3pt}c@{\hskip3pt}|@{\hskip3pt}c@{\hskip3pt}|@{\hskip3pt}c@{\hskip3pt}|@{\hskip3pt}c@{\hskip3pt}|@{\hskip3pt}c@{\hskip3pt}|@{\hskip3pt}c@{\hskip3pt}|@{\hskip3pt}c@{\hskip3pt}|@{\hskip3pt}c@{\hskip3pt}| }
  \hline
     seed&FM&MOx & $\Delta D$ &  $\Delta E$  & $\beta_{DMI}$  & $\eta$$_{DMI}$& ref \\ 
      nm &nm&nm &    mJ/m$^2$ &     MV/m   &  fJ/(V$\cdot$m) & 10$^{-12}$J$\cdot$m/C& \\  \hline
        Pt&Co&AlO$_x$/HfO$_x$&0.14-0.26&133&1100-2000&3.9-7.2&here\\ 
       3& 0.49&6-53&$\pm$0.2&&$\pm$700& &\\ \hline
        Ta&FeCoB &TaO$_x$/AlO$_x$/HfO$_x$& 0.1&  167&600 &3.2&\cite{Srivastava2018}*\\
      3& 0.65&1-10-50&  &&& &   \\ \hline
    Ta&FeCoB &TaO$_x$/AlO$_x$/HfO$_x$&0.105& 667& 158 &0.61&\cite{Srivastava2018}\\ 
       3& 0.65&1-10-50& & && &     \\ \hline
     Au&Fe &MgO/SiO$_2$ &$4\times10^{-5}$& 12.5&3.2 &1.72&\cite{citeulike:13767084}\\ 
       50& 20&10-270& & & &&   \\ \hline
     Pt&Fe& MgO&0.06&800&75&2.65&\cite{Zhang2018}*\\ 
        4& 2&367& & & &&  \\ \hline
     Ta/Pt&Co/Pd&MgO/HfO$_x$&0.038&577&67  &0.38&\cite{Koyama2018}\\ \
      2.6-2.4& 0.78-0.4&2-50& & & &&  \\ \hline
              V/Fe&Co& MgO/SiO$_2$&$1.8\times 10^{-3}$ & 18&100  &5.5&\cite{Suwardy2019}\\  
       20-20& 0.14 &5-50&  & & && \\ \hline
         V/Fe&Co& MgO/SiO$_2$&$1.2\times 10^{-3}$ & 18&65  &14.5&\cite{Suwardy2018} \\ 
       20-20&0.26&5-50&&&& & \\ \hline
           Pt&Co& MgO&& &26 &0.6-1.76 &\cite{Yang2018}\\ 
       & 0.6 &&  & & && \\ \hline

   \end{tabular} 
    \label{tab:DMI}
 }
 
   \caption{Summary of DMI variations under EF. The long time scale studies (hours our days)  have a star $*$ in the last column. }
\end{table}

To discuss the DMI variations under EF, we use the voltage-control DMI (VCDMI) coefficient \cite{Yang2018,Srivastava2018}), defined as  $\beta_{DMI}=\Delta D/\Delta E$ (in J/(V$\cdot$m), with $\Delta E=\Delta V/t_{Ox}$, where  $\Delta V$ is the voltage variation and  $t_{Ox}$ the dielectric tickness.  The coefficient $\beta_{\mathit{DMI}}$ is plotted versus $A_{\mathit{ex}}$ in Fig.  \ref{fig:BDMI}. It is in the range of $\beta_{DMI}= $ 1100 - 2000 $\pm$ 700 fJ/(V$\cdot$m).  In order to compare this result with previous works we give a summary of the different values in Table~1. 
Very spread $\beta_{DMI}$ values between 3.2 and 2000 fJ/(V$\cdot$m) are obtained. However, these values are not directly comparable, as  in addition to different materials and material quality, different thickness ranges of the ferromagnetic layer and different dielectric oxide layers have been used. To obtain comparable values of the  (VCDMI) coefficient, we define a new normalized coefficient  $\eta$$_{DMI}$,  which we introduce as the variation of  surface DMI constant $D_s$ per surface charge provided at the interface of the ferromagnet: $\eta$$_{DMI}= \Delta D_s/(\epsilon _e\Delta E)$, where  $\epsilon_e$ is the effective permittivity equal to $\epsilon_e=\epsilon_0\epsilon_r$ in the case of a single dielectric layer, where $\epsilon_0$ is the permittivity of  vaccum and $\epsilon_r$ the relative permittivity of the dielectric. This new normalized  $\eta$$_{DMI}$ coefficient takes into account the fact that the DMI is of interfacial origin, and consequently it is the variation of surface DMI constant  $D_s$ which should be compared.  In addition,  as high-k oxides have been used in order to allow smaller EF for a similar effect on DMI  we  also have to take into account the effective permittivity of the (possibly multilayered) oxide and compare the effect for a given induced surface charge density and not a given electric field (assuming that the EF effect is induced by electron displacement). 
After performing this renormalization of the EF changes on DMI, we can now compare the $\eta$$_{DMI}$ values corresponding to different studies from several teams. The obtained values lie within the range 0.6 to 14.5 $\times 10^{-12}$ J$\cdot$m/C. We first see that the two studies with long time scale  measurements (made by Brillouin light Spectroscopy (BLS) \cite{Srivastava2018,Zhang2018}) provide much larger $\eta$$_{DMI}$. This is  potentially due to the ion migration contribution from these long measurements. Long (days or hours) and short time (seconds or minutes) measurements present a factor of 5 ratio in  $\eta$$_{DMI}$ \cite{Srivastava2018}.
Our present result obtained within minute-timescale in ultrathin Co gives intermediate  $\eta$$_{DMI}$. 
For the case of the theoretical paper by Yang et al., we have reported two values for the $\eta$$_{DMI}$, considering or not that the EF calculated corresponds to an applied EF or takes into account the effect of $\epsilon_r$ in the oxide, which is still debated in the community, in particular due to the underestimation of the dielectric permittivity of MgO. These two values are thus giving ranges for theoretical $\eta$$_{DMI}$ in the Pt/Co/MgO system.
The relatively small dispersion in the extracted  $\eta$$_{DMI}$ values is noticeable, as all these studies have been performed in different teams with samples with different ferromagnetic materials, seed layers and oxides, grown with different techniques (magnetron sputtering or molecular beam epitaxy),  measured with different methods (BLS, DW asymmetric motion under in-plane magnetic field, analysis of the labyrinthine domain structures, excitation of magneto-static surface spin-waves) with EF ranging from 18 to 800 MV/m. This indicates that the underlying physics is similar for all these samples despite the very different $\beta_{DMI}$ proposed. 
In order to carefully compare interface effect and  optimize the EF effect on DMI, one should thus perform calculation of this renormalized, thickness and oxide-independent $\eta$$_{DMI}$.

\section*{\label{sec:con}Conclusion}
In conclusion, we were able to deduce a strong EF variation of the magnetic anisotropy energy ($\beta_{\mathit{PMA}}$ = 562 $\pm$ 102 fJ/(V$\cdot$m)) and of the $DMI$ term ($\beta_{\mathit{DMI}}$ $\sim$ 1100 - 2000 $\pm$ 700 fJ/(V$\cdot$m)). The strong observed variation of domain size $L_{eq}$ is a result of the combination of variations of every magnetic parameter. The strong $\beta$ coefficients we were able to deduce for PMA and DMI energies are the result of significantly increased EF effects for measurement temperatures close to $T_c$.

\section*{\label{sec:con}Acknowledgements}
 The authors thank M. Chshiev and A. Thiaville for fruitful discussions and C. Mahony for his contribution to the editing of the manuscript. This work was supported by the French ANR via Contract No. ELECSPIN ANR-16-CE24-0018), the DARPA
TEE program through Grant MIPR No. HR0011831554 and the EUR Grant NanoX No. ANR-17-EURE-0009 in the framework of the “Programme des Investissements d’Avenir”.

%
%\bibliographystyle{elsarticle-num-names} 
% \bibliography{stripes1}

\end{document}